\documentclass[aps,prl,twocolumn,showpacs]{revtex4}
\usepackage{amsmath}
\usepackage{epsfig}
\usepackage{subfigure}
\usepackage{graphicx}

\begin{document}

\title{Magnetic circular dichroism in GaMnAs: (no)
 evidence for an impurity band}

\author{Marko Turek$^{1, 2}$}
\email{marko.turek@physik.uni-regensburg.de}
\author{Jens Siewert$^1$}
\author{Jaroslav Fabian$^1$}
\affiliation{$^1$Institut f{\"u}r Theoretische Physik, Universit{\"a}t Regensburg,
             D-93040 Regensburg, Germany}
\affiliation{$^2$Fraunhofer Center for Silicon
Photovoltaics, W.-H\"{u}lse-Strasse 1, 06120 Halle (Saale), Germany  }

\begin{abstract}
Magneto-optical properties of the ferromagnetic semiconductor GaMnAs
are studied in a material specific multi-band tight-binding approach.
Two realistic models are compared: one has no impurity band while the other
shows an impurity band for low Mn concentrations. The calculated
magnetic circular dichroism (MCD) is positive for {\it both} models
proving that, unlike previously asserted, the observed positive MCD signal
is inconclusive as to the presence or absence of an impurity
band in GaMnAs. The positive MCD is due to the antiferromagnetic
p-d coupling and the transitions into the conduction band.
\end{abstract}


\maketitle

The ferromagnetic semiconductor GaMnAs
\cite{Ohno98, Jungwirth06a, Dietl08} is a fascinating material
not only because of its importance for technological applications
\cite{Fabian07}, but also because its understanding poses great
challenge. Indeed, understanding the electronic structure of GaMnAs
requires resolving complexities due to disorder, correlations, and
magnetism. Over the past years, an intense debate has sparked about
the nature of the carrier states (holes) \cite{Jungwirth07a}: Do the relevant states
reside in a GaAs-like valence band or in an Mn-induced impurity band? While the leading mean-field picture
\cite{Jungwirth06a} of GaMnAs relies on the valence-band scenario, also supported by
transport experiments \cite{Elsen07, Neumaier09}, recent optical experiments appear
to imply the existence of an impurity band \cite{Burch08}.

Ferromagnetism in GaMnAs is mediated by free holes
induced by Mn impurities \cite{Dietl00, DasSarma2003}. Experimentally,
magneto-optical studies of GaMnAs
\cite{Burch08,Szczytko96,Beschoten99,Chakarvorty07,Szczytko99,Ando08}
gave important insights into
the exchange mechanism between the holes and the local Mn moments \cite{Burch08}.
In particular, magnetic circular dichroism (MCD), which is the
relative absorption strength for the left and right circularly polarized
light, can give the signs and magnitudes of
the exchange coupling constants \cite{Oppeneer00,Ando00}. It is
now generally believed that the coupling between holes and the
local moments (p-d coupling) is antiferromagnetic.

The observed MCD signal around the fundamental absorption edge,
$E_{\rm gap} \approx 1.5 \; {\rm eV}$, is positive, which seems to imply
a ferromagnetic p-d coupling, in contrast to what is known
about II-Mn-VI diluted magnetic semiconductors \cite{Szczytko96,Lang05},
as well as about GaMnAs at very low doping levels, at which the material is
paramagnetic \cite{Schneider87}. Keeping the coupling antiferromagnetic,
the natural explanation, within the valence-band picture, appeared
to be a shift of the Fermi energy due to the Mn doping---Mn acts as
an acceptor---in the ferromagnetic regime. This so called Moss-Burstein
shift would then reverse the order in which the two circular polarizations of
light are absorbed \cite{Szczytko99,Szczytko01,Dietl01}.

On the other hand, recent MCD experiments have
been interpreted to imply the existence of an impurity band \cite{Ando08},
invalidating the Moss-Burstein shift picture. Such a view appears
consistent with the calculations of Ref. \onlinecite{Lang05},
which concluded that the positive MCD would arise in the valence-band
scenario only if the p-d coupling were ferromagnetic (this argument
counters the results of Ref. \cite{Hankiewicz04} which finds instead a positive
MCD in the valence-band model with Mn disorder in the Born approximation). Also,
recent tight-binding calculations for an embedded single Mn impurity in a
GaAs matrix \cite{Tang08} point to the existence of an impurity band. The
apparent picture coming from experiments \cite{Ando98, Chakarvorty07, Ando08, Burch08}
is that of a dominant and spectrally broad positive MCD signal from the
transitions to an impurity band.

Here we argue that the positive MCD signal has little to do with the
presence or absence of an impurity band in GaMnAs. The signal results
from the spin-resolved electronic level ordering in GaMnAs which
is present despite the strong disorder. The Fermi level shift
due to doping by Mn acceptors is strong enough to make the MCD signal positive,
regardless whether or not the Fermi level lies in the valence
band (Moss-Burstein shift) or in the impurity band, while holding
to the antiferromagnetic p-d coupling in {\it both} cases. The dominant
transitions are those involving the conduction band, not the impurity one.
We thus find most previous conclusions drawn from the positive MCD signal unfounded.
Our argument is based on large-scale tight-binding
calculations of MCD in disordered GaMnAs using two models. One
model reproduces the valence-band picture, the other shows an impurity
band. Both have antiferromagnetic p-d coupling, and both show a positive
MCD around the fundamental absorption edge, in agreement with experiment.

Our simulations use a material-specific microscopic
tight-binding approach. The
models are based on 16 $sp^3$ valence and conduction bands of
GaAs which are approximated throughout the entire Brillouin zone
to fit the experimentally determined band structure \cite{Chadi77a,Talwar82a}.
We use two different parameter
sets for the inclusion of Mn impurities into the models.
One of the models \cite{Masek07b} is characterized by an only slightly changed
host valence band and a shift of the Fermi energy according to
the number of holes that are added with the Mn. The other model \cite{Tang04a}
shows a strong shift of the host valence-band states into the gap leading
to the formation of an impurity band which starts to merge with the
valence band at Mn concentrations around $1\%$ \cite{Turek08a}. We present results on the
absorption of left and right circularly polarized light for
ferromagnetic bulk ${\rm Ga}_{1-x}{\rm Mn}_x{\rm As}$
at zero temperature and various Mn concentrations.
Within this framework we can treat the disorder effects
non-perturbatively.  While this approximation seems justified for
large concentrations of Mn impurities \cite{Jungwirth07a},  $x \gtrsim 1\%$, an
explicit inclusion of the carrier-carrier interactions
(computationally infeasible) may give  quantitative
corrections of our findings \cite{Dietl08a}.

As already stated, we investigate two different
sets of tight-binding parameterizations to study in
detail how an impurity band affects the MCD results.
These two models were thoroughly analyzed in
Ref.~\cite{Turek08a} concerning the density of states, the
size of the band gap, the position of the Fermi energy,
localization properties of holes, optical effective masses,
and mean free paths. The first model, which we call model A,
was suggested by Ma{\v s}ek \cite{Masek07b}. This model,
derived from a first-principles approach, describes
GaMnAs qualitatively very similar to what is expected from a p-doped
GaAs \cite{Turek08a}. Its main characteristic is the inclusion of
additional holes and therefore a corresponding
shift of the Fermi energy into the host valence band.
There is no formation of an impurity band within this model.

The second model, which we call model B, was suggested by
Tang and Flatt{\'e} \cite{Tang04a}. Within this model the Mn impurities
are described by a modified on-site potential and a spin-dependent
potential at the four nearest As neighbors. The two relevant parameters are
adjusted such that the experimental binding energy of $0.11~{\rm eV}$
of the Mn impurity is recovered. Qualitatively, this model not only
shifts the Fermi energy due to the additional holes but it also affects
the positions of the energy levels of the host material rather
strongly \cite{Turek08a}.
Applying this model to disordered systems
with many Mn impurities leads to the formation of an
isolated impurity band for
Mn concentrations $x \lesssim 1\%$. At higher concentrations the
impurity band starts to merge with the host valence band.
Hence, the comparison of these two different parameterizations, models
A and B, allows us to draw conclusions on how the existence of an
impurity band influences the MCD signals that are seen in the experiments.

\begin{figure}[t]
\centerline{\psfig{file=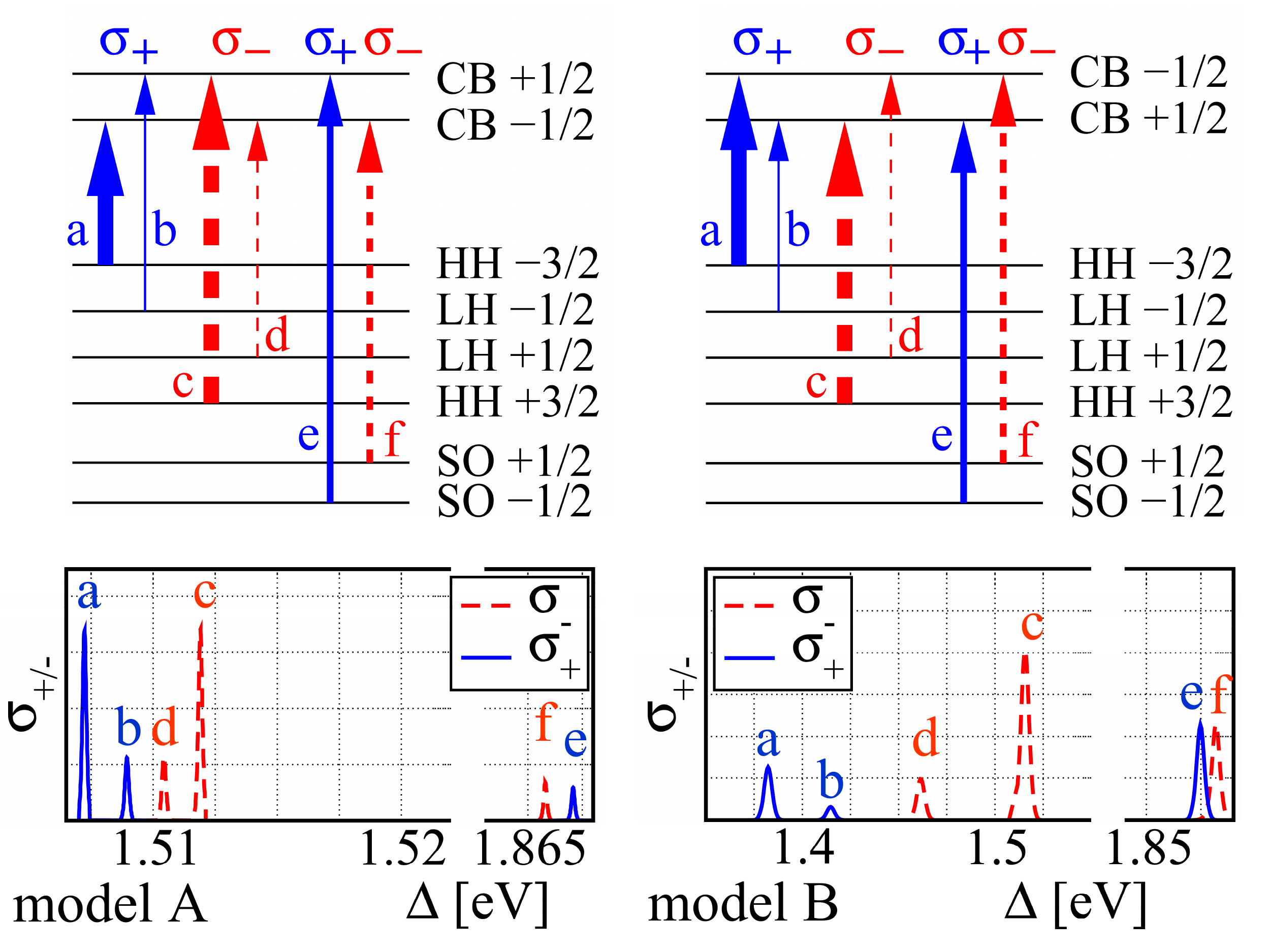,width=\linewidth}}
\caption{\label{fig_SingleMn_All}
 (Color online) Left panel: model A, right panel, model B. Upper panel:
   Schematic level ordering deduced from absorption data. Lower panel:
      Absorption of right (solid line) and
          left circularly polarized (dashed line) light. The system size is 2000 atoms with one Ga atom
	       replaced by Mn. The absorption rates are evaluated at the $\Gamma$ point in the
	             super-lattice Brillouin zone. Levels are labeled according to their
		     carrier character in pure GaAs: conduction band (CB), heavy and light-hole bands (HH, LH), and the spin-orbit
		     split-off band (SO). Magnetic quantum numbers are indicated. Since substitutional Mn is an acceptor, the highest valence band
		            level (`HH -3/2') is not occupied so that the transitions `a' are suppressed.
			           }
				          \end{figure}

We analyze the MCD signal by evaluating the
diagonal and off-diagonal elements of the optical absorption matrix.
The matrix elements $\sigma_{\alpha \beta}$
can be written in terms of the eigenstates
$|n\rangle$ and eigenenergies $E_n$ as \cite{Czycholl04}
\begin{equation}
 \label{eq:conductivity}
 \sigma_{\alpha \beta} (\Delta, E_F) = \frac{{\rm i} e^2 \hbar}{m^2 \Omega}
 \sum\limits_{n n'} \frac{f_n - f_{n'}}{\Delta_{n' n}}
 \frac{p^\alpha_{n n'} p^\beta_{n' n}}{\Delta + {\rm i} \delta - \Delta_{n' n}},
\end{equation}
with the system volume $\Omega$, the Fermi function $f_n = f_n(E_F)$,
energy difference $\Delta_{n' n} \equiv E_{n'} - E_{n}$, and momentum matrix elements
$p^\alpha_{n' n} \equiv \langle n' | \hat p^\alpha | n \rangle$.
The absorption $\sigma_\pm$
for right (left) polarized light can be obtained by replacing
$p \to p^{\pm} \equiv p^x \pm {\rm i} p^y$ in Eq.~(\ref{eq:conductivity}).
The MCD signal, for a constant index of refraction (or poor conductors), is
\begin{equation}
MCD = (\sigma_- - \sigma_+)/(\sigma_- + \sigma_+).
\end{equation}
In order to numerically evaluate the conductivity (\ref{eq:conductivity}) we
calculate the eigenenergies and the matrix elements of the momentum
operator using a multi-band tight-binding approach \cite{Turek08a}. This approach
is applied to periodically repeated finite size super-cells containing up to 2000
atoms. The resulting conductivity
is furthermore averaged over several disorder configurations.
Therefore, our approach goes beyond the single impurity calculation
presented on model B in Ref.~\cite{Tang08} as we explicitly include
disorder averaging effects. In our calculations, for numerical reasons,
we use a smearing temperature corresponding to $1$ meV in the Fermi functions
and a peak width $\delta = 5 \; {\rm meV}$. The integration
over the super-lattice Brillouin zone is performed by summation over up to 2048
different $\vec k$ vectors \cite{Turek08a}.
As the systems under consideration are
disordered one cannot restrict this summation to the irreducible
part of the Brillouin zone. Furthermore, the finiteness of the systems
limits the Mn concentrations to $x\gtrsim 1\%$ in our present simulations.

For a better understanding of the numerical MCD results we first
analyze the effect of a single Mn impurity in a super-cell of
2000 atoms with periodic boundary conditions.
The absorption was evaluated at the $\Gamma$ point, i.e.
$\vec k = 0$ in the super-lattice Brillouin zone.
Due to the magnetic moment of the Mn impurity the
spin degeneracies of the conduction s- and the valence p-bands are lifted.
The order in which the levels appear can be concluded from the absorption
peaks of $\sigma_\pm$ at $\Delta \approx E_{\rm gap}\approx 1.5 \; {\rm eV}$. We show
the numerical results in Fig.~\ref{fig_SingleMn_All}. The Fermi energy was chosen to be
(artificially) above the highest impurity level, to show all relevant optical transitions.
The absorption peaks then correspond to the six possible transitions between the six
heavy hole, light hole, and spin orbit split-off bands to the
two conduction bands \cite{Ando00}. Other transitions are suppressed
because of the selection rules for the momentum matrix elements.
Among the allowed transitions, `a' and `c' are the strongest ones
due to their larger momentum matrix elements. From the position of each
$\sigma_+$ and $\sigma_-$ peak one can uniquely identify
the involved bands. The extracted ordering
of the levels is shown schematically in the upper panel of
Fig.~\ref{fig_SingleMn_All}.

As a Mn impurity also acts as an acceptor, the highest impurity
level, that is the levels labeled `HH -3/2' in
Fig.~\ref{fig_SingleMn_All}, is in fact unoccupied. This
means that the actual Fermi energy lies
just below this highest impurity level.
Therefore, all transitions starting from this
level are suppressed. For the absorption
this means that the peak labeled `a' is not observable at
zero temperature.

Figure~\ref{fig_SingleMn_All} points to  common features and differences between the two models.
The order of the first four peaks, `a' through `d', is the same
for both models implying the same ordering of the valence band
states. The spin of the Mn core 3d electrons in our simulation points
into the $-z$ direction, that is, anti-parallel to the propagation direction
of the light. This leads to an acceptor state with spin-down
character implying antiferromagnetic p-d exchange coupling for the holes,
for both models.

Besides this common feature there are two major differences between model
A and model B.
First, the magnitude of the level splitting due to the magnetic impurity
is very different. For example, model A results in a splitting of
the four heavy and light hole bands of $\approx 4 \; {\rm meV}$ while in
model B this splitting is $\approx 0.13 \; {\rm eV}$. This difference
in the splittings is associated with the fact that model B explicitly
reproduces the bound state energies of a single Mn impurity at
$\approx 0.11 \; {\rm eV}$ in the host gap by a strong deformation of
the host valence band \cite{Tang04a}. The second major difference between the
two models is the different order of the two conduction bands.
In model A the spin-up band has a higher energy while in model B
it is the spin-down band. Hence we find that model A is based on
a ferromagnetic s-d coupling, consistent with the prevalent view,
while model B produces antiferromagnetic s-d coupling.
Concerning the absorption data this is reflected in the fact
that at energies $\Delta \approx E_{\rm gap} + \Delta_{\rm SO} \approx 1.8 \; {\rm eV}$
for model A the $\sigma_-$ peak comes first while for model B it is
the $\sigma_+$ peak. This specific in the
ordering of the conduction band does not alter the calculated MCD signal qualitatively.

\begin{figure}
\centerline{\psfig{file=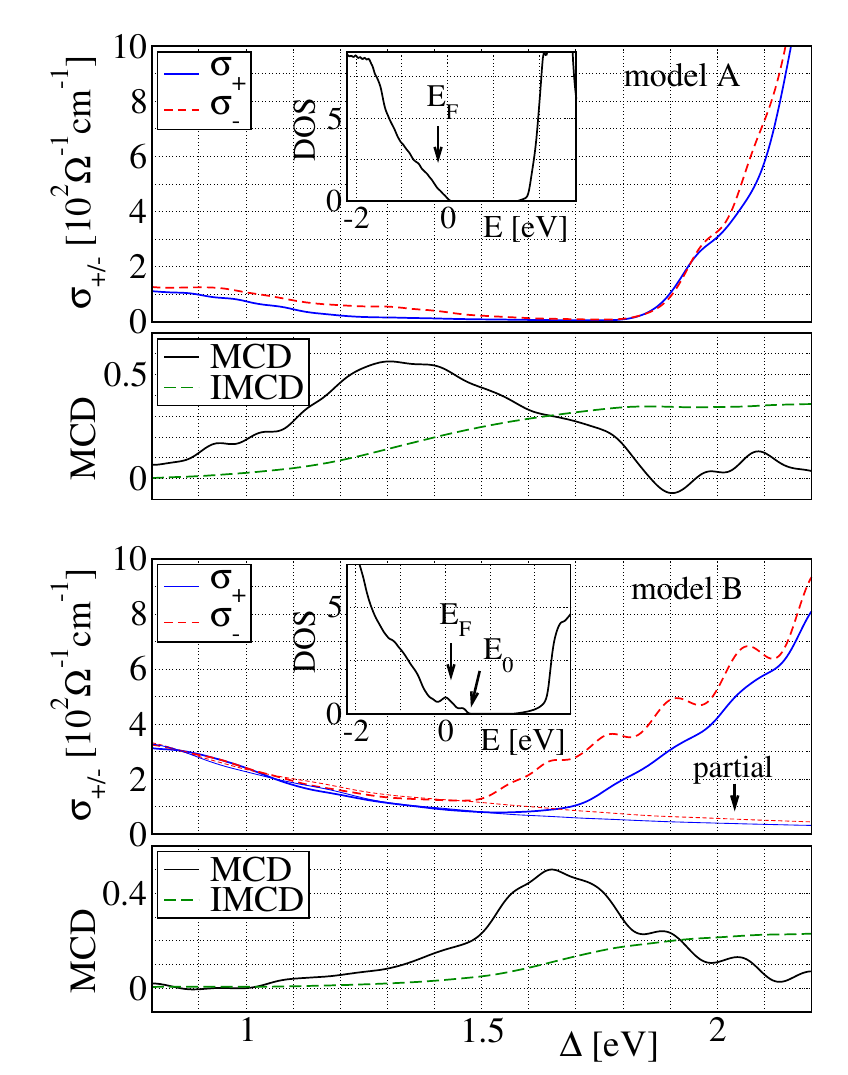,width=0.9\linewidth}}
\caption{\label{fig_2percMn_MCD}
 (Color online) Calculated absorption of right (solid line) and
 left (dashed line) circularly polarized light for model A (upper graph) and
 model B (lower graph). The accompanying thin lines
 in model B are the partial conductivities excluding the transitions to the
 conduction band. The system size is
 432 atoms with 2 \% Ga atoms
 replaced by Mn. The absorption rates are evaluated using 2048 $\vec k$ vectors (256 for
 the partial conductivities) in
 the entire Brillouin zone; disorder averaging was over 5 configurations,
 and a running average over $25$ meV was applied.
 From the absorption the MCD signal and the integrated MCD (IMCD) are determined.
 The insets show the density of states with the position of the Fermi
 energy (model A: $E_F=-0.23\;{\rm eV}$; model B: $E_F=0.13\;{\rm eV}$).
 For model B, the energy $E_0$ marks the top of the impurity band,
 $E_0=0.6\;{\rm eV}$.
 }
\end{figure}

How are the above effects of a single Mn impurity reflected in the
absorption of the left and right circularly polarized light in
{\it strongly} disordered samples with many impurities?
The simulation results for a Mn concentration $x=2\%$
are shown in Fig.~\ref{fig_2percMn_MCD}.
For this Mn concentration model B still shows an impurity band, which is attached
to the host valence band, while model A does not exhibit an impurity band
at all as can be seen in the insets of Fig.~\ref{fig_2percMn_MCD}.
The Fermi energy in model A lies within the host valence band
whereas for model B it lies within the impurity band.
For both models we find that the $\sigma_-$ absorption sets in at lower energies
$\Delta$ implying a positive MCD signal. This is understandable
as the strong $\sigma_+$ transitions at low $\Delta \approx E_{\rm gap}$
are suppressed as the highest valence and impurity band states
with spin down character are not occupied. Comparing with
recent experiments we find very good qualitative agreement for the
MCD and IMCD signals \cite{Beschoten99, Chakarvorty07}.
A precise quantitative agreement cannot be expected from a
tight-binding model \cite{Harrison}.

To test the
hypothesis put forward in \cite{Ando08}, that the positive MCD is due
to the transitions to the impurity band, we have calculated
the partial conductivities by excluding the conduction band states in
Eq. \ref{eq:conductivity} for model B; see Fig. ~\ref{fig_2percMn_MCD}.
These conductivities (and their differences) are much smaller than
the complete ones, giving no support to the hypothesis.
Instead, the positive MCD around the absorption edge is due to the
transitions to the conduction band.

\begin{figure}[t]
\resizebox{0.45\textwidth}{!}{\includegraphics{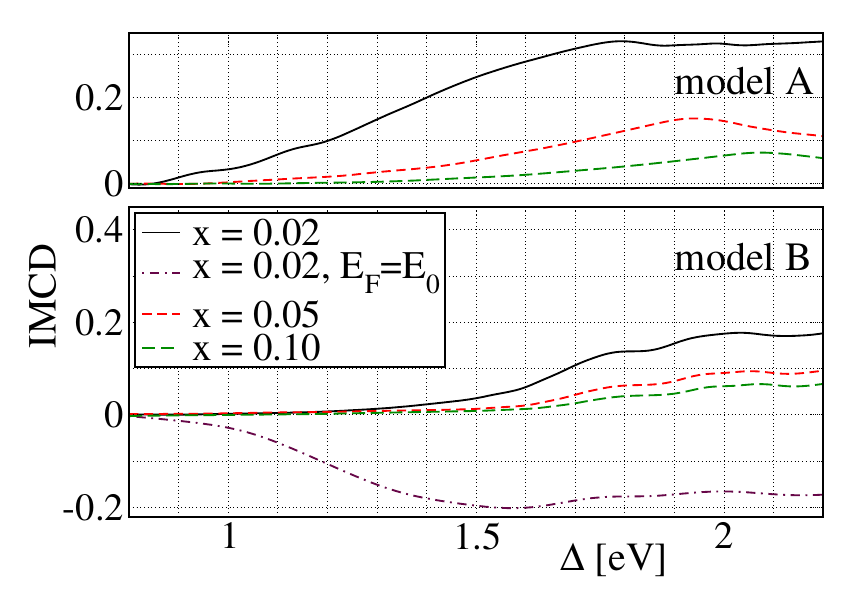}}
\caption{\label{fig_imcd_vs_conc}
 (Color online) Top panel: model A, lower panel: model B.
 Integrated MCD signal for different Mn concentrations
 $x=2 \%$ (solid line), $x=5 \%$ (short dashed), and $x=10 \%$ (long dashed).
 The system size is 432 atoms; the absorption rates were evaluated using 256 $\vec k$ vectors,
 and disorder averaged over 20 configurations.
 For model B we also show  the integrated MCD (IMCD) signal for
 Mn concentration $x=2 \%$ with the Fermi energy
 shifted to the top of the impurity band, $E_F=E_0$ (dashed dotted line).
}
\end{figure}

As in Ref.~\onlinecite{Chakarvorty07},
we show the IMCD signal for various Mn concentrations in
Fig.~\ref{fig_imcd_vs_conc}. With increasing Mn concentration the
magnitude of the IMCD signal is decreasing as the total absorption
becomes stronger. However, the  shape of the signal
is rather robust for all investigated Mn concentrations. In particular,
we do not observe a change in the sign of the MCD signal within
the considered range of concentrations $x$ from 2 \% to 10 \%.

What would the MCD signal be for a fully compensated GaMnAs?
It is particularly interesting to see what the MCD were if the
Fermi level were at the top of the impurity band in model B, as this reflects the
ordering of the impurity band levels and the deformation of the ordering
due to disorder. Our results are shown in Fig.~\ref{fig_imcd_vs_conc}
for $x=2$\%  ($E_F=E_0$). The MCD is negative as the
$\sigma_+$ transitions are now allowed, dominating the signal
in accord with the single-impurity picture in Fig. \ref{fig_SingleMn_All}.

In summary, we find that the positive MCD signal in GaMnAs
(a) is inconclusive as to the presence or absence of an impurity band,
(b) implies antiferromagnetic p-d coupling also for the valence-band scenario,
and (c) originates from the transitions to the conduction band.

We thank D.~Weiss and C.~Back for helpful discussions. This work was
supported by the DFG SFB 689.


\end{document}